\documentclass[twocolumn,superscriptaddress,showpacs,preprintnumbers,prl]{revtex4}

\usepackage{graphicx}

\newcommand{\twid}{\sim}

\newcommand{\gap}{\hspace{.4in}}

\newcommand{\gt}{\rightarrow}

\newcommand{\period}{\ \ .}
\newcommand{\comma}{\ ,\ }

\newcommand{\const}{\mbox{const.}\ }

\newcommand{\lsim}{\,\stackrel{<}{\scriptstyle \sim}\,}
\newcommand{\gsim}{\,\stackrel{>}{\scriptstyle \sim}\,}

\newcommand{\ignore}[1]{}

{\begin{tabular}{#1}}%
{\end{tabular}}

{\end{list}}

\newenvironment{benlistdefault}{%
\begin{list}{}{\setlength{\labelwidth}{1.0in}%
                \setlength{\labelsep}{.5in} 
                \setlength{\leftmargin}{1.5in} }%
}%
{\end{list}}

{\ \\ XXXXX \hfill XXXXX #1 XXXXX\begin{equation}\label{#1}}%
{\end{equation}}

{\ \\ XXXXX \hfill XXXXX #1 XXXXX\begin{eqnarray}\label{#1}}%
{\end{eqnarray}}

\newenvironment{bequation}[1]%
{\begin{equation}\label{#1}}%
{\end{equation}}

\newenvironment{beqnarray}[1]%
{\begin{eqnarray}\label{#1}}%
{\end{eqnarray}}

\newcommand{\ie}{i.\,e.~}

{\begin{center} \section*{#1} \end{center}}

{\end{figure}} 

{\end{figure}}

{\end{figure}}

\newenvironment{eq}[1]%
{\begin{bequation}{#1}}{\end{bequation}}

{\begin{beqnarray}{#1}}{\end{beqnarray}}

\def\eqref#1{(\ref{#1})}

\newcommand{\taus}{\tau_s}

\newcommand{\smax}{s_{\rm max}}

\newcommand{\tauN}{\tau_N}

\newcommand{\phisurf}{\phi_{\rm surf}}

\newcommand{\phistar}{\phi^{*}}
\newcommand{\Zsurf}{Z_{\rm surf}}
\newcommand{\Zbulk}{Z_{\rm bulk}}

\newcommand{\kT}{k_B T}

\newcommand{\Gammabound}{\Gamma_{\rm bound}}
\newcommand{\gammabound}{\gamma_{\rm bound}}
\newcommand{\Gammadot}{\dot{\Gamma}}

\newcommand{\Rtotal}{{\cal R}_{\rm total}}

\newcommand{\Omegat}{\Omega_t}
\newcommand{\Omegatdot}{\dot{\Omega}_t}

\newcommand{\tsatchem}{t_{\rm sat}^{\rm chem}}
\newcommand{\tsatphys}{t_{\rm sat}^{\rm phys}}
\newcommand{\taubulk}{\tau_{\rm bulk}} 
\newcommand{\tadsorb}{t_{\rm adsorb}} 

\newcommand{\RF}{R_F}

\newcommand{\ta}{t_a}

\newcommand{\ncont}{n_{\rm cont}}

\newcommand{\rhosuper}{\rho_{\rm super}}
\newcommand{\Gammaboundinf}{\Gamma_{\rm bound}^{\infty}}
\newcommand{\Gammainf}{\Gamma^{\infty}}

\newcommand{\lsep}{l_{\rm sep}}

\newcommand{\nth}{^{\rm th}}

\begin{document}



\title{Irreversibility and Polymer Adsorption}

\author{Ben O'Shaughnessy}

\affiliation{Department of Chemical Engineering, Columbia University, New York, NY 10027}

\author{Dimitrios Vavylonis$^{1,}$}

\affiliation{Department of Physics, Columbia University, New York, NY 10027}



\begin{abstract}

Physisorption or chemisorption from dilute polymer solutions often
entails irreversible polymer-surface bonding. We present a theory of
the non-equilibrium layers which result. While the density profile and
loop distribution are the same as for equilibrium layers, the final
layer comprises a tightly bound inner part plus an outer part whose
chains make only $fN$ surface contacts where $N$ is chain length. The
contact fractions $f$ follow a broad distribution, $P(f)\twid
f^{-4/5}$, in rather close agreement with strong physisorption
experiments [H. M. Schneider et al, Langmuir {\bf 12}, 994 (1996)].

\end{abstract}


\ignore{
PACS numbers:
  \begin{benlistdefault}
    \item [82.35.-x]
   (Polymers: properties; reactions; polymerization)
    \item [05.40.-a]
   (Fluctuation phenomena, random processes, noise, and Brownian Motion)
     \item [68.08.-p]
   (Liquid-solid interfaces)
\end{benlistdefault}
} 

\maketitle


The validity of the laws of equilibrium statistical mechanics hinges
on ergodicity, the ability of a system to freely explore its phase
space \cite{ma:book}.  Many real processes, however, involve
irreversible microscopic events such as strong physical or chemical
bonding which invalidate ergodicity. Equilibrium then becomes
inaccessible and Boltzmann's entropy hypothesis is no longer applicable
to calculate observables.  Instead, the kinetics must be followed from
their very beginning: the accessible region of phase space is
progressively diminished as successive irreversible events freeze in
an ever-increasing number of constraints. The state of the system at
some time depends on the pocket of phase space to which it has become
confined.

\begin{figure}
\includegraphics[width=8.6cm]{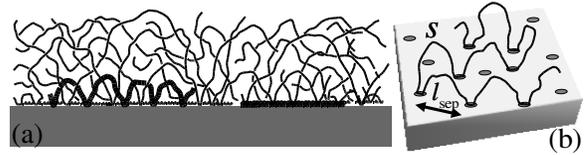}
\caption{\label{layer} (a) Final irreversible layer
structure.  Chains  highlighted in bold: one belongs to the
inner flattened layer ($\omega N$ surface contacts) the other to the
outer layer ($f N \ll N$ contacts, loop size
$s\approx \ncont/f$). (b) Late stage chain adsorption as surface
approaches saturation and free supersites (clusters of $\ncont$ empty
sites) become dilute. Chains cannot completely zip down. The minimum
loop size $s$ just connects two nearest neighbor supersites separated
by $\lsep$, \ie $as^{3/5}=\lsep$.  }
\end{figure}

The adsorption of high molecular weight polymers onto surfaces by its
very nature frequently involves this kind of irreversibility (see
fig. \ref{layer}).  When an attractive surface contacts even a very
dilute polymer solution there is a powerful tendency for dense polymer
layers to develop
\cite{fleer:pol_iface_book,gennes:polads_theory:combo_aip} because
sticking energies per chain increase in proportion to the number of
monomer units, $N$.  This effect is exploited in many technologies
such as coating, lubrication, and adhesion.  When the monomer sticking
advantage $\epsilon$ exceeds $\kT$, available experimental evidence
indicates that relaxation times become so large that the physisorption
processes are effectively irreversible
\cite{granick:irr_physi_bimodal:combo_aip}.  This is a common
situation.  Many polymer species attach through strong
hydrogen bonds \cite{joestenschaad:book:h_bonding} ($\epsilon\gsim
4\kT$) to silicon, glass or metal surfaces in their naturally oxidized
states \cite{granick:irr_physi_bimodal:combo_aip}, while DNA and
proteins adhere tenaciously to a large variety of materials through
hydrogen bonds, bare charge interactions or hydrophobic forces
\cite{hlady:protein_dna_ads:combo_aip}. In such situations layer
structure is no longer determined by the laws of equilibrium
statistical mechanics.  The extreme example arises in {\em
chemisorption}
\cite{lenk:chemi_exp:combo_aip,shaffer:chak_chemi:combo_aip} where
covalent surface-polymer bonds develop irreversibly as in applications
such as polymer-fiber welding in fiber-reinforced thermoplastics and
colloid stabilization by chemical grafting of
polymers\cite{chemiphysiletter:applications}.  Generally, applications
prefer the strongest and most enduring interfaces possible and
irreversible effects are probably the rule rather than the exception.

Our aim in this letter is to understand the effect of irreversibility
on the structure of adsorbed polymer layers (see fig. \ref{layer}).
Polymer adsorption phenomena are a major focus of polymer science, and
though a few theoretical and numerical works have addressed
irreversibility
\cite{shaffer:chak_chemi:combo_aip,barford:irr_ads:combo_aip,shaffer:strong_ads_heteropolymers}
the reversible case and the equilibrium layers which result are far
better understood
\cite{fleer:pol_iface_book,gennes:polads_theory:combo_aip}.  Theory
\cite{gennes:polads_theory:combo_aip}, consistent with a number of
experiments \cite{lee:polads_expts:etal_aip}, predicts each
adsorbed chain in the equilibrium layer has sequences of surface-bound
monomers (trains) interspersed with portions extending away from the
surface (tails and loops of size $s$). For good solvents the loop
distribution $\Omega(s)\twid s^{-11/5}$ and net layer density profile
$c(z) \twid z^{-4/3}$ are universal.  Equilibrium and ergodicity imply
every chain is statistically identical. For example, for large $N$ the
fraction $f$ of units which are surface-bound is the same for all
chains to within small fluctuations and is no different to the overall
bound fraction, $f = \Gammabound/\Gamma$. Here $\Gamma$ is the total
adsorbed polymer mass per unit area and $\Gammabound$ the
surface-bound part.

\begin{figure}
\includegraphics[width=8.3cm]{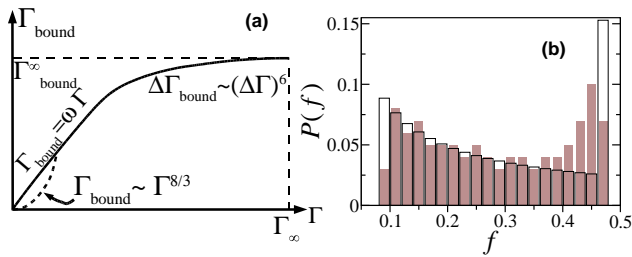}
\caption{\label{granick}
(a) Predicted adsorbed polymer mass $\Gamma$ versus surface-bound part
$\Gammabound$.  For chemisorption, 
$\Gammabound\twid \Gamma^{8/3}$ initially.  (b) Frequency histograms for
fraction of bound mass, $f$. Experiment (grey) from ref. \onlinecite{granick:irr_physi_bimodal:combo_aip}. Theory
(empty), from predicted distribution $P(f)\twid f^{-4/5}$ with
$f_{\rm max} < f \lsim \omega$, where values for $f_{\rm max} = 0.9$
and $\omega = 0.47$ were taken from
ref. \onlinecite{granick:irr_physi_bimodal:combo_aip}. 
}
\end{figure}

How are these universal features modified when the adsorption is
irreversible? This question was explored in a series of ingenious
experiments by the workers of refs.
\onlinecite{granick:irr_physi_bimodal:combo_aip} who monitored
polymethylmethacrylate (PMMA) adsorption from dilute solution onto
oxidized silicon via hydrogen bonding with $\epsilon\approx
4\kT$. Measuring infrared absorption and dichroism, they monitored
both $\Gamma(t)$ and $\Gammabound(t)$ as they evolved in time and
showed that early arriving chains had much higher $f$ values than late
arrivers and these $f$ values were frozen in for ever.  They
modeled \cite{granick:irr_physi_bimodal:combo_aip} this in terms of
a picture where early arrivers lie flat and late arrivers having fewer
available surface spots to adsorb onto are extended.  The experimental
$f$ values of the asymptotic layer followed a broad distribution,
shown in fig. \ref{granick}(b). This succinctly quantifies the
essential non-ergodic characteristic of these non-equilibrium layers:
there are now infinitely many classes of chains, each class with its
own particular statistics.

\begin{figure}
\includegraphics[width=7.5cm]{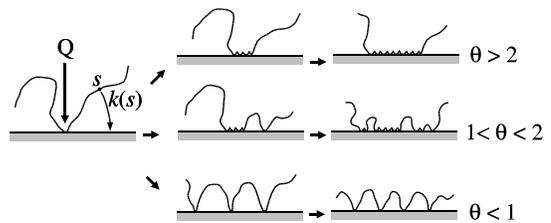}
\caption{\label{collapse}
Chain adsorption commences with formation of an initial
monomer-surface bond. For chemisorption, the reaction rate thereafter
for the $s\nth$ monomer from this graft point is $k(s)\twid
s^{-\theta}$. Three modes of subsequent chain adsorption are
theoretically possible: zipping ($\theta>2$); accelerated zipping,
where occasional big loops nucleate new zipping centers
($1<\theta<2$); and uniform collapse ($\theta<1$). Chemisorption from
dilute solution is accelerated zipping ($\theta=8/5$).
}
\end{figure}

In the following an initially empty surface contacting a dilute
polymer solution with good solvent is considered.  We will calculate
the kinetics of layer formation, $\Gammabound(t)$ and $\Gamma(t)$, and
the distributions of $f$ values and loop sizes in the evolving and
final layer.  The two cases of irreversible physisorption and
chemisorption must be carefully distinguished. Define $Q$ as the
``reaction'' rate between a monomer and the surface, given this
monomer contacts the surface (see fig. \ref{collapse}).  For
physisorption, the attachment of a monomer is virtually instantaneous
on reaching the surface so the effective value is diffusion-limited,
$Q\approx 1/\ta \approx 10^{10}$ sec$^{-1}$ typically, where $\ta$ is
monomer relaxation time. Chemisorption processes are much slower, with
typical values \cite{ben:chemiphysilettet_combo} $10^{-2} \lsim Q
\lsim 10^2$ sec$^{-1}$.  Consider a chain which, having diffused from
bulk to surface, has just made its first attachment, \ie just one
monomer is irreversibly bonded to the surface (see
fig. \ref{collapse}). We first treat the case of chemisorption, where
the subsequent attachment of the remaining monomers is a process
lasting seconds to hours and is thus experimentally accessible (all
$N$ monomers are assumed functionalized).

(1) {\em Early stages: single chain adsorption and surface
saturation.} How does this chain adsorb down onto the surface?  This
depends on the exponent $\theta$ governing the surface reaction rate
$k(s)$ for the s$\nth$ monomer measured from the initial graft point
(see fig. \ref{collapse})
                                                   \begin{eq}{dog}
k(s) \approx Q\, \Zsurf(s,N) / \Zsurf(N) \approx Q/s^{\theta}
\comma\
					(s\ll N) \period
                                                                \end{eq}
Here $\Zsurf(N)$ and $\Zsurf(s,N)$ are the chain partition functions
given one and two surface attachments, respectively.  Slow
chemisorption allows sufficient time for chains to explore all
configurations given the current constraints frozen in by earlier
reactions.  Eq. \eqref{dog} states the reaction rate is
proportional to the fraction of the grafted chain's configurations for
which the s$\nth$ monomer contacts the
surface\cite{shaffer:chak_chemi:combo_aip}. Now in cases where
$\theta>2$, the total reaction rate $\Rtotal \approx \int_1^N ds\,
k(s)$ is dominated by $s$ of order unity, \ie monomers near to the
first attached monomer will attach next. Thus, the chain {\em zips}
down from the initial graft point. In contrast, for systems where
$\theta<1$ the upper limit dominates $\Rtotal$, \ie a distant monomer
will react next; this implies a much more homogeneous chain {\em
collapse} mechanism (see fig. \ref{collapse}).

The present situation is a self-avoiding polymer at a repulsive wall
(we consider pure chemisorption, \ie we assume a free energy advantage
for solvent to contact the wall.)  It turns out this case is
intermediate between zipping and collapse. By relating $\theta$ to
other polymer exponents at hard walls \cite{duplantier:networks} we
obtained the exact relation
                                                \begin{eq}{cat}
		\theta = 1 + \nu 
                                                                \end{eq}
where $\nu\approx 3/5$ is the Flory exponent
\cite{duplantier:networks} determining the polymer bulk coil size
$\RF=a N^\nu$ in good solvent ($a$ is monomer size). Thus $1<\theta<2$
and $\Rtotal$ is dominated by its lower integration limit. We call
this case {\em accelerated zipping} (see fig. \ref{collapse}). Zipping
from the original graft point is accompanied by the occasional
grafting of a distant monomer producing a loop of size $s$, say. This
occurs after time $\taus \approx 1/ \int_s^N ds'\, k(s')\approx Q^{-1}
s^{3/5}$.  Each such new graft point nucleates further zipping,
enhancing the effective zipping speed. Hence the entire chain adsorbs
in a time $\tadsorb = \tauN \approx Q^{-1} N^{3/5}$, since by this
time even the biggest loops have come down. Note this is much less
than the pure zipping time $\approx Q^{-1} N$. Thus pure zipping
must have been short-circuited by large loop adsorption events before
it could have completed its course.

During this accelerated zipping down, a characteristic (unnormalized)
loop distribution $\Omegat(s)$ develops and the number of
surface-bound monomers $\gammabound(t)$ grows from 1 to order $N$. We
calculated these quantities by solving the detailed loop
kinetics\cite{chemiphysi:note_loopkinetics}.  These are rather
complex, and here we present more accessible scaling arguments which
reproduce the same results.  Let us postulate that after time $t$ the
only relevant loop scale is the largest to have come down, $\smax
\approx (Qt)^{5/3}$, \ie $\Omegat(s) \approx (\smax/s)^{\alpha}/\smax$
for $s\ll \smax$.  Assuming $\alpha>1$, the total number of loops
$L(t) \approx \smax^{\alpha-1}$ is dominated by small loops of order
unity. Writing $\gammabound(t) = N (t/\tauN)^{\beta}$ we demand this
be independent of $N$ for $t\ll\tauN$ (imagine sending the chain size
to infinity; this would not affect the accelerated zipping propagating
outwards from the initial graft point).  This determines $\beta =
5/3$. Finally, since there are $L(t)$ nucleating points for further
zipping, $ d\gammabound/dt \twid L$, \ie $\gammabound\twid Lt$
which fixes $\alpha=7/5$.

We now sum over all chains which attached up to time $t$. The
entropic disadvantage to touch the surface reduces the monomer volume
fraction at the surface from the far field bulk value $\phi$,
$\phisurf = r \phi$ where the ratio of surface to bulk chain partition
functions $r\equiv \Zsurf(N) /\Zbulk(N) = 1/N$ was calculated in
ref. \onlinecite{duplantier:networks}. Then with
\cite{chemiphysi:note_interiordominate} $a^2 d\Gamma/dt = Q N
\phisurf$ and $\Gammabound = \gammabound (\Gamma/N)$ we have
                                                \begin{eq}{bee}
 \Gamma(t) a^2 = \phi Q t\comma\ \ \  \Gammabound(t) a^2 = \phi N^{3/5}
(t/\tauN)^{8/3} \comma  
                                                                \end{eq}
describing the early chemisorption layer for $ t<\tadsorb=Q^{-1}
N^{3/5}$. The loop structure of the partially collapsed chains is
$\Omegat(s) \twid s^{-7/5}$ with maximum size $\smax= (Qt)^{5/3}$.
This first phase may be long lived; e.g. for $Q^{-1} = 1$ sec.,
$N=10^3$ then $\tauN \approx 20$ mins. This becomes many hours for
smaller $Q$ values which are common.

By  time $\tadsorb$ zipping is complete and each chain is completely
flattened onto the surface with fraction of adsorbed monomers
$f=\omega$.  The species-dependent constant $\omega$ is of order unity
and reflects steric constraints preventing every monomer from 
actually touching the surface.  In practice, we expect broadening of $f$ values
about $\omega$ due to strong fluctuations, typical of multiplicative random
processes characterizing irreversibility. For longer times each new
chain zips down and $\Gammabound(t) = \omega \Gamma(t)$ with $\Gamma$
given by eq. \eqref{bee}. This proceeds until $\tsatchem \approx 1/(Q
\phi)$ when the surface is virtually saturated with a near-monolayer
of flattened chains\cite{chemiphysi:note_independentadsorption}.

Consider now physisorption in its early stages. After attachment of
its first monomer, the collapse of a single chain into a flattened
structure now occurs as rapidly as monomers can diffuse a distance of
order $\RF$, possibly accelerated by the attachments themselves. Thus
we expect the collapse time \cite{shaffer:strong_ads_heteropolymers}
to be at least as small as the bulk coil relaxation time $\taubulk$
(of order microseconds). Hence the collapse itself is
probably experimentally unobservable, at least with the techniques of
ref. \onlinecite{granick:irr_physi_bimodal:combo_aip}.  What is
important is that in dilute solutions chains
collapse into flattened configurations without hindrance from
others. Moreover, we find that the probability a chain arriving from
the bulk makes at least one bond before diffusing away is
essentially unity even for a nearly-saturated surface.  It follows
that the attachment of chains is diffusion-controlled for essentially
all times, $a^2 \Gamma(t)\approx (\phi/a)(Dt)^{1/2}$ where $D$ is
center of gravity diffusivity.  As for chemisorption, $\Gammabound =
\omega \Gamma$ and  adsorption produces a virtual monolayer of
flattened chains. Surface saturation effects onset after time
$\tsatphys= \taubulk (\phistar/\phi)^2 N^{2/5}$.

(2) {\em Late stages: the tenuously attached outer layer.} Both
chemisorption and physisorption processes fill the surface with
completely collapsed chains, albeit in very different timescales
$\tsatchem$ and $\tsatphys$.  By this stage the distribution of
surface-bound fractions is sharply peaked at $f=\omega$.  However, as
saturation is approached free surface sites become scarce and
late-arriving chains can no longer zip down completely. Suppose each
chain-surface adhesion point consists in $\ncont$ attached monomers.
The precise value of $\ncont$ is sterically determined and is expected
to be strongly species dependent.  Then the surface density of free
``supersites'' (unoccupied surface patches large enough to accommodate
$\ncont$ monomers) is $\rhosuper \approx \Delta\Gammabound/\ncont$
where $\Delta\Gammabound \equiv \Gammaboundinf - \Gammabound$ is the
density of available surface sites and $\Gammaboundinf$ is the
asymptotic density of bound monomers.  Now as the surface approaches
saturation so the density of supersites becomes small, $\rhosuper \ll
1/(\ncont a^2)$, and their mean separation $\lsep \approx
\rhosuper^{-1/2}$ becomes so large that a late-arriving chain cannot
find contiguous supersites to complete its accelerated zipping
down. The minimum loop size $s$ which can come down is that just large
enough to connect two free supersites, \ie $a s^{3/5} = \lsep$ whence
$s=(\ncont/a^2 \Delta\Gammabound)^{5/6}$.  Thus the final adsorbed
state of chains arriving at this stage (see fig. \ref{layer}(b))
consists of trains of $\ncont$ monomers separated by loops of
order $s$ units. For these chains $\partial
\Delta\Gammabound/\partial\Delta \Gamma = f \approx \ncont/s$ for
large $s$, where $\Delta\Gamma$ is the deviation from the asymptotic
coverage $\Gammainf$. Integrating this process up to saturation,
                                                \begin{eq}{pepper}
a^2 \Delta\Gammabound = \ncont (a^2 \Delta\Gamma/6)^6\comma\ 
		P(f) = Af^{-4/5}\ \ 
                                                                \end{eq}
where $f \ll 1$ and $A$ is a constant of order unity
\cite{chemiphysi:note_pfderivation}.  Adding this broad
distribution of $f$ values to the peak centered at $f=\omega$ from the
early stages gives the total distribution, shown in
fig. \ref{granick}(b). It agrees rather closely with the experimental
one of ref. \onlinecite{granick:irr_physi_bimodal:combo_aip} shown in the
same figure.  The predicted $\Gammabound(\Gamma)$ profile (see
fig. \ref{granick}(a)) is also very close to the measured
profile\cite{granick:irr_physi_bimodal:combo_aip}.

Eq. \eqref{pepper} describes a tenuously attached outer
layer (small $f$ values) formed by late arriving chains, adding to the
dense flattened layer formed at earlier times. The loop distribution
of this diffuse outer layer is obtained from $s\Omega(s) ds/\Gammainf
= P(f) df$ whence
                                                \begin{eq}{brown}
\Omega (s) \approx a^{-2} s^{-11/5}\comma\gap c(z) \twid z^{-4/3}
                                                                \end{eq}
where the density profile followed from $c=\Omega s ds/dz$ evaluated
at $z=a s^{3/5}$.

Finally, the kinetics of the total and bound coverages during the late
stages are modified by saturation effects.  For chemisorption the rate
of attachment is directly proportional to the density of available
surface sites, $\Gammadot \twid \Delta\Gammabound\twid\
(\Delta\Gamma)^6$ so $\Delta\Gamma\twid t^{-1/5}$ and
$\Delta\Gammabound\twid t^{-6/5}$. In the physisorption case as
discussed diffusion-control always pertains, $\Gamma\twid t^{1/2}$,
and thus the bound fraction saturates as $\Delta\Gammabound \twid
(1-\const t/\tsatphys)^6$.

In conclusion, we found that irreversible adsorption of polymer chains
leads to final non-equilibrium layers exhibiting both similarities and
profound differences compared to their equilibrium counterparts. The
layer is a sum of a surface monolayer plus a diffuse outer part of
thickness of order the bulk coil size with density profile $c(z)\twid
z^{-4/3}$ and loop size distribution $\Omega(s) \twid
s^{-11/5}$. Interestingly, these features are identical to those
predicted for equilibrium layers, including the precise exponent
values. Prefactors are different, however, and we anticipate different
values for physisorption and chemisorption. To determine these
necessitates accounting for topological constraints and fluctuations
in empty surface site densities and other quantities, effects absent
from our model.  Note that although we did not explicitly treat
excluded volume interactions between an adsorbing chain and those
previously adsorbed, we expect these to be unimportant because an
empty site is correlated with a reduced surface loop density at that
location.

What is very different about irreversible layers is that individual
chains in the layer are not statistically identical: a given chain
either belongs to the surface bound part and has order $N$ surface
contacts, or else the diffuse outer part. In the latter case the
number of contacts, $fN$, is generally much less than $N$ and its loop
distribution is almost monodisperse with loop size $s\twid 1/f$.  In
equilibrium layers there is just one class of chain; parts of each
chain lie bound to the surface, other parts extend into the outer
layer and its loop distribution is the same as the layer's.  In
contrast, for irreversible layers there are an infinite number of
classes, each with its own $f$ value. The weighting for different
values is universal, $P(f)\twid f^{-4/5}$ for small $f$.
Practically, these differences have important implications for
the physical properties of irreversible layers; for example, the outer
layer is much more fragile than the protected inner flattened layer.
From a fundamental point of view these systems provide a measurable
example of how irreversible events progressively diminish the
available phase space volume and modify the entropy algorithm. For an
equilibrium layer with $\epsilon>kT$ this gives
\cite{gennes:polads_theory:combo_aip} $F \approx F_{\rm trans} +
F_{\rm osm} + F_{\rm train}$ for the free energy. Here $F_{\rm trans}
= - \kT \int ds \Omega(s) \ln [a^2 \Omega(s)]$ derives from loop
translational entropy, $F_{\rm train} = E_{\rm train} - TS_{\rm
train}$ is the contribution from trains and $F_{\rm osm}$ is the
osmotic part due to the solvent-swollen loops in the outer part of the
brush. By comparison, for the non-equilibrium layers both trains and
loops are immobilized on the surface, $F_{\rm trans} = S_{\rm
train}=0$. The free energy is thus increased, $F \approx F_{\rm osm} +
E_{\rm surf}$. Its modified structure is expected, for example, to
profoundly modify the interaction between polymer-covered surfaces as
compared to the equilibrium case where the rearrangement of chains on
the surfaces leads to characteristic force profiles
\cite{gennes:polads_theory:combo_aip}.

This work was supported by the National Science Foundation, grant
no. DMR-9816374.



\begin{thebibliography}{10}
\bibitem{chemiphysi:note_loopkinetics}
The loop kinetics are $\Omegatdot = \int_s^N ds' \Omegat(s') k(s|s') - \int_0^s
  ds' \Omegat(s) k(s'|s)$ where $k(s|s')$, the rate $s'$-loops generate 2 loops
  $s$ and $s'-s$, has the small $s$ behavior of eq. \eqref{dog}, $k(s|s') \gt
  k(s) \twid s^{-\theta}$ for $s\ll s'$.
\bibitem{chemiphysi:note_interiordominate}
Although chain ends are more likely to touch the surface than a typical
  interior monomer, we find the latter dominate since there are order $N$ of
  them.
\bibitem{chemiphysi:note_independentadsorption}
The surface density of chains at $t=\tadsorb$ is $\Gamma(\tadsorb)/N
  \approx(\phi/\phistar)/ \RF^2 $ where\cite{gennes:polads_theory:combo_aip}
  $\phistar = N^{-4/5}$ is the overlap threshold. Thus for dilute conditions
  ($\phi<\phistar$) a chain zips down flat before others arrive to interfere.
\bibitem{chemiphysi:note_pfderivation}
The distribution was obtained from
  $P(f)=1/(\Gammainf[\Delta\Gammabound^{''}]_{\Delta\Gammabound'=f})$ where
  prime denotes differentiation with respect to $\Delta\Gamma$. The prefactor
  is $A\approx 6/(5 a^2\Gammainf [\ncont]^{1/5})$.
\end{thebibliography}

\end{document}